\documentclass[useAMS,usenatbib]{mn2e}
\usepackage{graphicx}

\title[The flux ratio of the OIII $\lambda$$\lambda$5007, 4959 lines in
AGN]{The flux ratio of the [OIII] $\lambda$$\lambda$5007, 4959
lines in AGN: Comparison with theoretical calculations}

 \author[M.S. Dimitrijevi\'c, L.\v C. Popovi\'c, J. Kova\v cevi\'c, M. Da\v ci\'c
and D. Ili\'c] { M. S.
Dimitrijevi\'c$^{1}$\thanks{E-mail: mdimitrijevic@aob.bg.ac.yu
(MSD); lpopovic@aob.bg.ac.yu (L\v CP); jkovacevic@aob.bg.ac.yu
(JK); mdacic@aob.bg.ac.yu (MD); dilic@matf.bg.ac.yu (DI)}, L. \v
C. Popovi\'c$^{1}$, J. Kova\v cevi\'c$^{1,2}$, M. Da\v ci\'c$^{1}$ and D. Ili\'c$^{2}$\\
$^{1}$Astronomical  Observatory,  Volgina  7, 11160  Belgrade
74, Serbia \\
$^{2}$ Department of Astronomy, Faculty of Mathematics, University
of Belgrade, Studentski trg 16, 11000 Belgrade, Serbia}

\date{Released 2006 Xxxxx XX }

\pagerange{\pageref{firstpage}--\pageref{lastpage}} \pubyear{2006}

\begin{document}

\label{firstpage}

\maketitle

\begin{abstract}
{By taking into account relativistic corrections to the magnetic
dipole operator, the theoretical [OIII] 5006.843/4958.511 line
intensity ratio of 2.98 is obtained. In order to check this new
value using AGN spectra we present the measurements of the flux
ratio of the [OIII] $\lambda\lambda$4959, 5007 emission lines for a
sample of 62 AGN, obtained from the Sloan Digital Sky Survey (SDSS)
Database and from published observations. We select only high
signal-to-noise ratio spectra for which the line shapes of the
[OIII] $\lambda\lambda$4959,5007 lines are the same. We obtained an
averaged flux ratio of $2.993 \pm 0.014$, which is in a good
agreement with the theoretical one. }
\end{abstract}


\section{Introduction}

The forbidden [OIII] $\lambda$4958.911 \AA \ $(2s^22p^2 {\ }^1D_2 -
2s^22p^2 {\ }^3P_1)$ and $\lambda$5006.843 \AA \ $(2s^22p^2 {\
}^1D_2 - 2s^22p^2 {\ }^3P_2)$ spectral lines are among the most
prominent emission lines, not only in the spectra of photoionized
nebulae, but also in the spectra of photoionized gas around Active
Galactic Nuclei (AGN) due to the relatively high abundance of doubly
charged oxygen ions. These lines are typical for AGN and originate
from the ionized Narrow Line Region (NLR) gas surrounding the
accreting super massive black hole in the centre \citep[see][]{b13}.
It should be emphasized that since they are located in the centre of
the visible band they are very often observed in spectra of HII
regions, photoionized nebulae and AGN. Because of observational and
physical circumstances this pair of lines are suitable to: (a) test
observationally the accuracy of theoretical calculations from atomic
theory; (b) check the linearity of the detectors in use; (c)
eventually test the assumptions on the target physics under extreme
circumstances (optical thickness effects).

These two spectral lines are the result of magnetic dipole
transitions with a small contribution of electric quadrupole
radiation. The elaborate theoretical work of \citet{b9} provided the
[OIII] 5006.843/4958.911 intensity ratio value of 2.89. \cite{b16}
checked this result by using photoionized gaseous nebulae spectra,
where these lines can be observed with a very high signal-to-noise
ratio. They found a small, but well-established, difference of 4-9
per cent between observations and theory. Namely \citet{b15} deduced
an intensity ratio of $3.03 \pm 0.03$, while measurements of
\citet{b10} provided a value of $3.17 \pm 0.04$, and that of
\citet{b12} the value of $3.00 \pm 0.08$.

In order to improve the agreement between observations and theory,
\citet{b16} took into account the relativistic corrections to the
magnetic dipole operator, demonstrating that they affect the
transition probabilities for the [OIII] $\lambda$5006.843 \AA\ and
$\lambda$4958.911 \AA\ \ lines. They obtained the A- value ratio of
3.01, implying a line intensity ratio of 2.98, which is only two
percent or less different from the values \citet{b15} and
\citet{b12} obtained from gaseous nebulae spectra, and 6 per cent
different from the value \citet{b10} obtained from the central
region of starburst galaxy Tololo 1924-416 (H II regions).
\cite{b10} found a spatial variation of the forbidden line
parameters, where the ratio is found to vary from $2.63\pm0.15$ to
$3.33\pm0.07$ along the slit (see their table 3 and fig. 10) with a
mean value of $3.17\pm0.04$, which significantly deviates from
theoretical ratios \citep{b16}. The authors mentioned that such
results might be caused by the detector's nonlinearity.

\citet{b16} also underlined the necessity to additionally check
their theoretical improvement of the line intensity ratio by the
corresponding observation in photoionized gaseous nebulae spectra.
However, due to instrumentational improvements the accuracy and
resolution of observed spectra has increased to the point where we
can now use AGN spectra for such purposes.

The [OIII] $\lambda\lambda$4959,5007 lines  originate in the NLR of
an AGN, the region with  conditions  which differ from those in
photoionized gaseous nebulae: i) the emission comes from a spatially
very extended region, so that one can expect quite different
physical and kinematical conditions  in different parts of a
NLR\footnote{Note here that the [OIII] lines observed in AGN very
often show a blue asymmetry and substructure in shapes (see e.g.
\cite{le06})}, ii) the dust on large spatial scales can result in
orientation-dependent effects on NLR line fluxes.  However, one can
expect that the forbidden-line emission is isotropic since
self-absorption in narrow lines is negligible. Therefore, due to the
significance of this pair of lines, it is important to check whether
their flux ratio is in agreement with theory. Only then can it be
reliably used for different checks of theoretical assumptions
concerning the physics of the NLR and photoionized gaseous nebulae.

The aim of this paper is to check, with the help of a large sample
of  AGN spectra, the improved theoretical value of the [OIII]
5006.843/4958.511 line intensity ratio \citep{b16}. Additionally,
 we want to investigate the usability of the AGN [OIII]
5006.843/4958.511 emission line flux ratio for checks of various
theoretical assumptions and for checks of linearity of detectors in
use. In order to do so we will measure the considered flux ratio of
the [OIII] lines in a large sample of AGN. Moreover, we will derive
some explicitly or implicitly given flux ratios of the [OIII]
$\lambda\lambda$5007, 4959 lines in the existing literature
\citep{b12a,b0,b6}, obtained for galaxies and quasars in order to
compare them with the \citet{b16} [OIII] 5006.843/4958.511 line
intensity ratio and with our results.

\begin{figure*}
\includegraphics[width=0.32\textwidth]{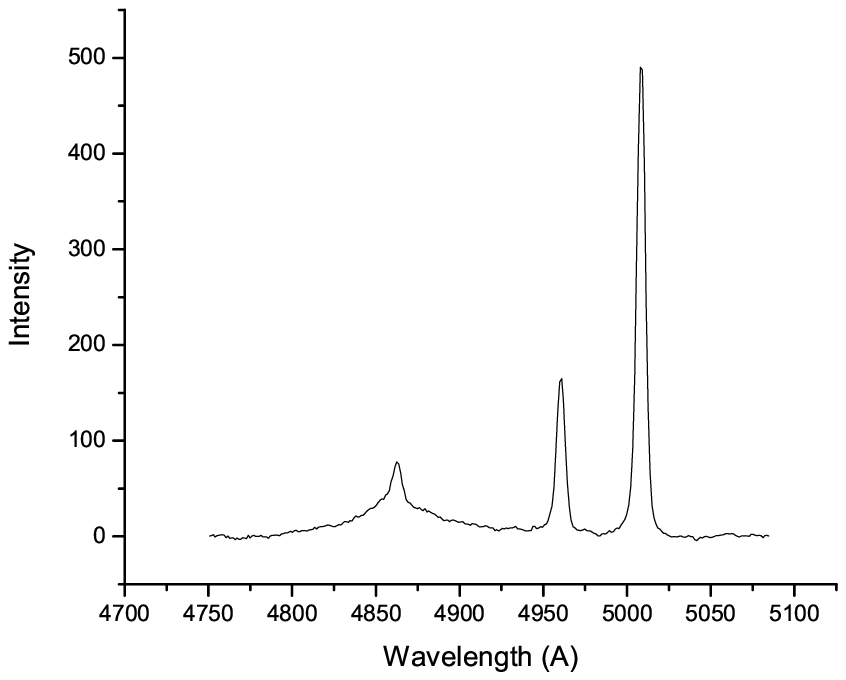}
\includegraphics[width=0.32\textwidth]{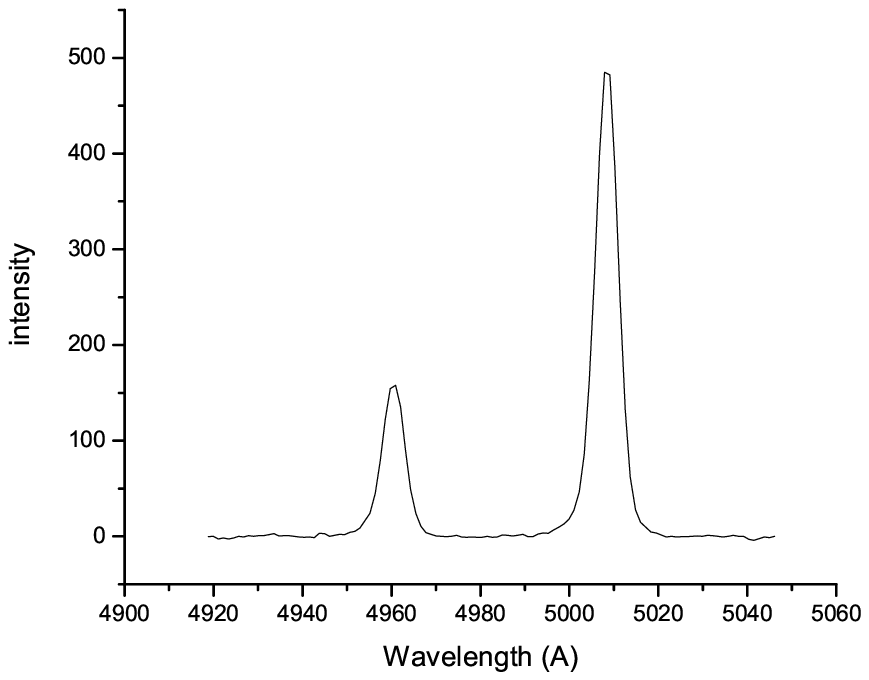}
\includegraphics[width=0.32\textwidth]{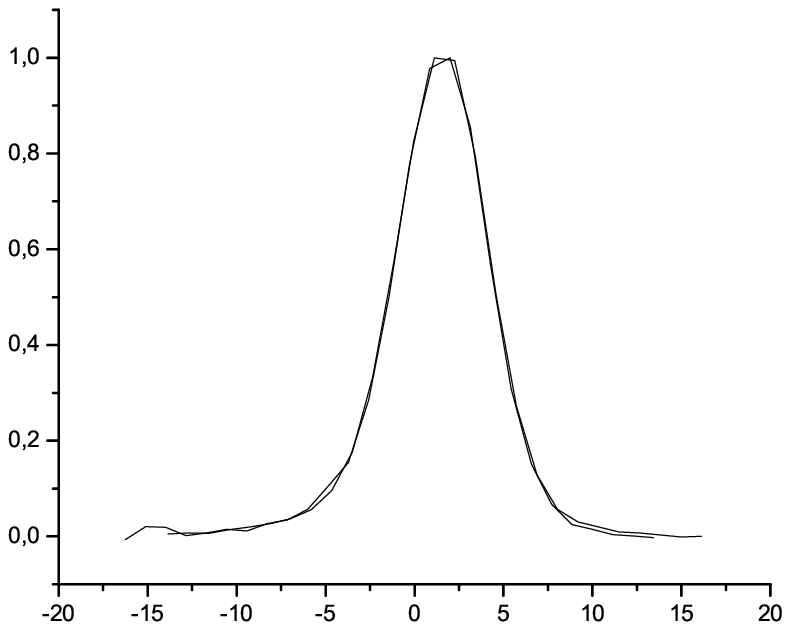}
\caption{Example of the selected spectrum (SDSS
J082308.29+42252000.00) with the same shapes of the [OIII] $\lambda$
5007 and $\lambda$4959  lines. Left - observed spectrum, in the
middle - lines without continuum and contaminating emission, right -
the profile of $\lambda$4959 line scaled to the profile of
$\lambda$5007 line. }
\end{figure*}

\begin{figure*}
\includegraphics[width=0.32\textwidth]{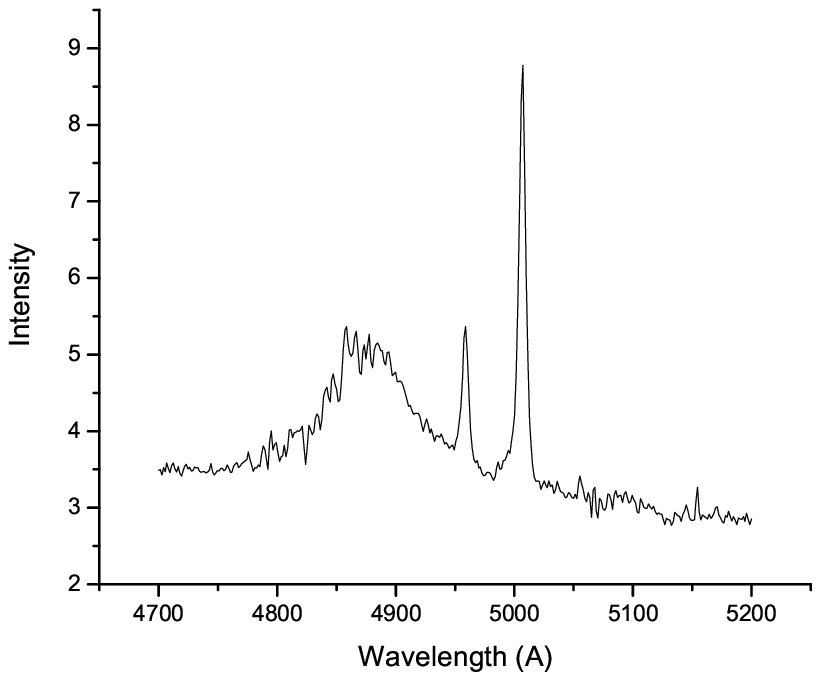}
\includegraphics[width=0.32\textwidth]{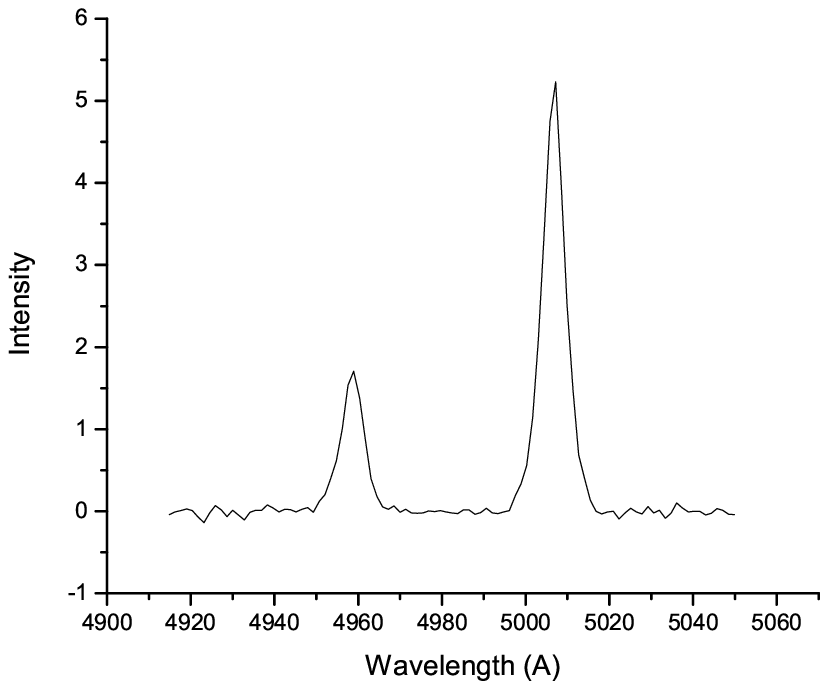}
\includegraphics[width=0.32\textwidth]{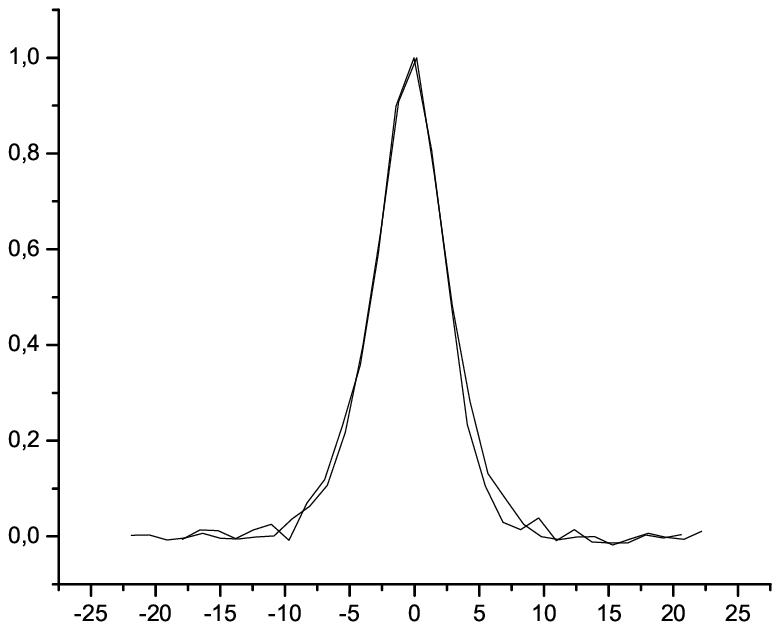} \caption{Example of the
 spectrum (PKS 2135-14) where the line shapes are slightly different in the line wings.}
 \end{figure*}
 \begin{figure*}
\includegraphics[width=0.32\textwidth]{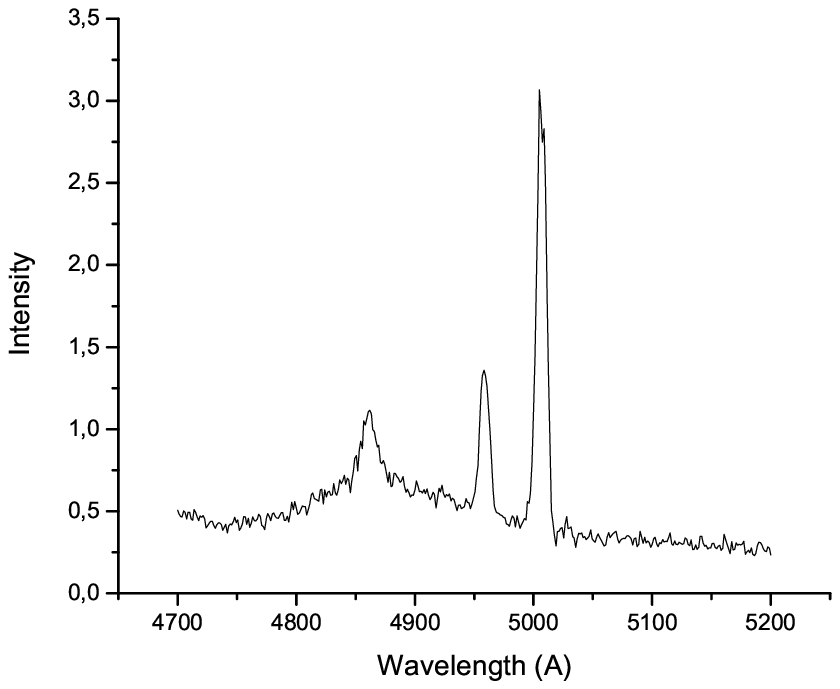}
\includegraphics[width=0.32\textwidth]{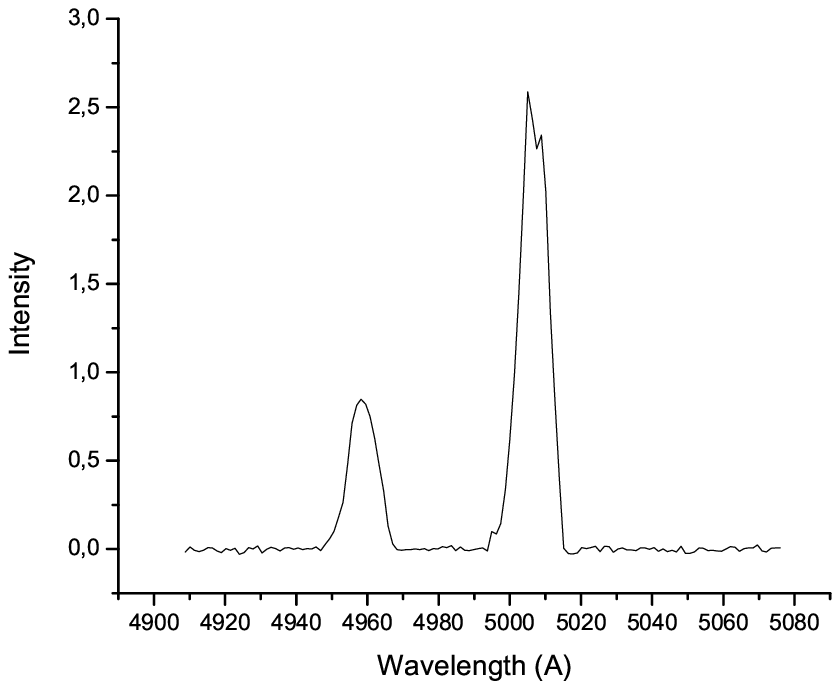}
\includegraphics[width=0.32\textwidth]{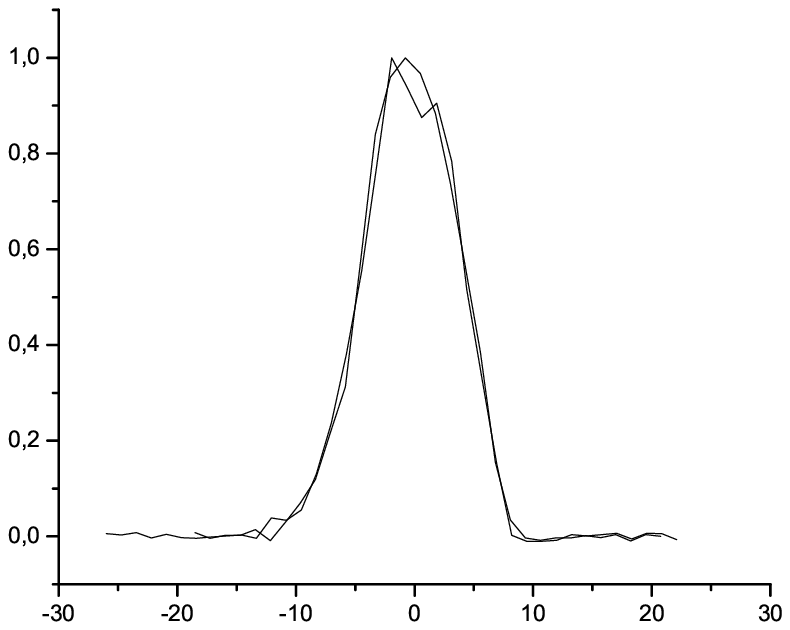} \caption{Example of the
 spectrum (PKS 2300-68) where the line shapes are slightly different in the central part.}
\end{figure*}

\section{The sample and measurements}

We selected our AGN sample, choosing the spectra with high
signal-to-noise ratio, from Data Release Four (DR4) of the SDSS
Database and from the observations described in the paper of
\citet{b18}.  The SDSS spectra cover the wavelength region 3800 \AA\
- 9200 \AA . It was shown that the flux-calibration is a few percent
on average, which is impressive for a fibre-fed spectrograph
\citep{t04}. \cite{t04} found that 1$\sigma$ error in the synthetic
colours is 5 per cent in g(4700\AA )-r(6200\AA ). The remaining
(small) flux-calibration  residuals are coherent on scales of 500
\AA , which has negligible effect on our flux measurements that are
obtained using an interval of less than 100 \AA . Consequently, we
exclude the effects of nonlinearity of the detector on the measured
[OIII] line ratio.

In our analysis, we first subtracted the continuum by using DIPSO
software package. In some spectra the $H_{\beta}$ and Fe II emission
lines, which contaminate the [OIII] $\lambda\lambda$4959, 5007,
lines, were subtracted.

In order to arrive at a clean sample we defined the following
selection criterion: when scaling the profile of the weaker
$\lambda$4959 emission line into the stronger $\lambda$5007 emission
line profile, the line profiles of the two lines in question should
differ insignificantly in the given spectrum (Fig. 1). This assures
that the measurements of flux ratios and of line intensity peak
ratios yield identical results. Examples of the maximal difference
in the line wings and in the central part of the profiles are shown
in Fig. 2 and Fig. 3, respectively. In these examples the observed
spectra, the lines without continuum, contaminating Fe II and
H$_{\beta}$ emission, and the profile of $\lambda$4959 scaled into
the profile of $\lambda$5007 are presented. Our initial sample of 62
AGN was selected using the criterion that the shapes of the both
lines are same or that the difference is negligible as shown in Figs
(1-3).

Following the criterion outlined above, from the initial sample of
62 AGN, we selected 56, then 40 and finally 34 spectra, discarding
spectra with slightly different [OIII] $\lambda\lambda$4959, 5007
line shapes (Figs. 2, 3). The final sample of 34 AGN has the best
matching of the [OIII] $\lambda\lambda$ 4959, 5007 line profiles
(Fig. 1). We  measured the flux ratio for each sample and we present
here histograms of the flux ratio values of the initial sample of 62
AGN (Fig. 4) and of the final sample of 34 AGN (Fig. 5).

\begin{table}\centering
\caption{The flux ratio $R$ of the [OIII] $\lambda\lambda$5007, 4959
lines of 62 AGN. The initial sample of 62 AGN is successively
reduced to 34 by discarding line profiles of lower quality.}
\begin{tabular}{@{}cc}
  \hline

  Number of AGN in the sample & $R$\\
  \hline
  62 & $2.992 \pm 0.014$ \\
  56 & $2.986 \pm 0.012$ \\
  40 & $2.994 \pm 0.014$ \\
  34 & $2.993 \pm 0.014$ \\
  \hline
\end{tabular}
\end{table}

\begin{table}
 \centering
  \caption{The flux ratio $R$ of the [OIII] $\lambda \lambda $4959, 5007 lines of
12 NLS1 galaxies, derived on the basis of \citet{b6}.}
  \begin{tabular}{@{}lc}
    \hline
    Object & $R$ \\
    \hline
    Mrk 705 &      3.00  \\
    Mrk 1239 &     2.90  \\
    Mrk 734 &      2.92 \\
    NGC 4748 &     2.99  \\
    Mrk 783 &      2.92  \\
    IRAS 13224-3809 &  2.94\\
    CTS J13.12 &   3.02 \\
    IRAS 15091-2107 &  2.93 \\
    Mrk 291 &      2.95  \\
    RXS J20002-5417 &   2.94 \\
    ESO 399-IG 20   & 3.04 \\
    Mrk 896 &      2.88 \\
    \hline
    The average value &   $2.953 \pm 0.014$\\
  \end{tabular}
\end{table}

\section{Results and discussion}

\subsection{The flux ratio of the [OIII] $\lambda\lambda$ 4959, 5007 lines
in spectra of active galaxies and quasars}

The [OIII] $\lambda$5006.843 \AA\ and $\lambda$4958.911 \AA\ both
originate from the same lower and slightly different upper energy
level and both have a negligible optical depth since the transitions
are strongly forbidden, therefore both may be scaled to exactly the
same emission-line profile. Moreover, \citet{b0} have shown that the
effect of differential reddening on the line splitting is of the
order of $10^{-8} \tau_{5007}$, i.e. negligible. If there are
multiple clouds that contribute to the emission, the observed
emission-line profiles are composed of the same mixture of
individual cloud complexes \citep{b0}.

Despite the fact that spectra of galaxies and quasars have not been
used to explicitly check the theoretical flux ratio of the [OIII]
$\lambda\lambda$5007, 4959 lines, there are examples where such
ratios were obtained as a by-product or could be derived from
published results \citep{b12a,b0,b6}.

For example, \citet{b12a} investigated the Seyfert 1.2 galaxy Mkn
79 with long-slit spectroscopy, using the intensity ratio
of the [OIII] $\lambda\lambda$4959, 5007 lines to check the
accuracy of the flux measurements along the slit. Their fig. 2
shows the measured ratios of [OIII] $\lambda\lambda$4959, 5007
lines plotted against position along slit. They report that the
ratio is very close to the value of 2.94, but due to the scatter
of results and error bars given in their figure, we excluded them
from present considerations.

\citet{b0} used these [OIII] lines of quasars with $0.16 < z <
0.80$, obtained from the SDSS Early Data release, to test whether
the fine-structure constant depends on cosmic time. As a by-product,
they found that the ratio of transition probabilities corresponding
to the [OIII] $\lambda\lambda$5007, 4959 lines is $2.99 \pm 0.02$.
We note that this result obtained from spectra of quasars is in
agreement with the theoretical value given by \citet{b16}.
\citet{b16} obtained a theoretical ratio of transition probabilities
of 3.01, which is within the given error bars of the observationally
derived value.

\citet{b6} measured the NLR emission-line flux ratios relative to
the flux of the $H{_{\beta}}$ line for 12 Narrow-Line Seyfert 1
(NLS1) galaxies. Using the [OIII] 4959/$H{_{\beta}}$ and [OIII]
5007/$H{_{\beta}}$ flux ratios given in \citet{b6}, we derived the
flux ratio of the [OIII] $\lambda$$\lambda$5007, 4959 lines for
these 12 NLS1 galaxies. The results are given in Table 2. The
average value for the ratio of observed fluxes corrected for
internal reddening is $2.953 \pm 0.014$, which also supports the
theoretical improvement of \citet{b16}.

\begin{figure}
\includegraphics[height=5cm]{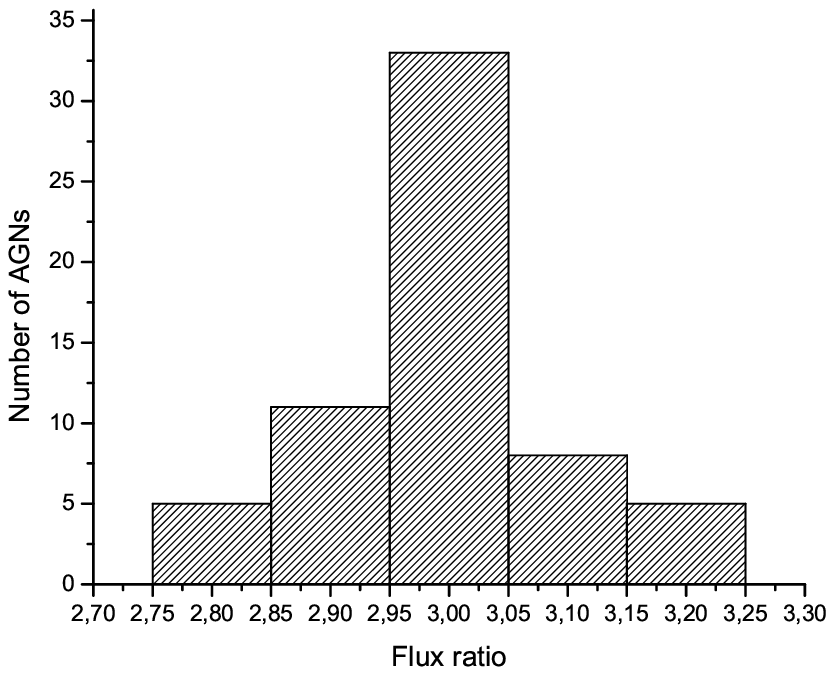}
\caption{Histogram showing the distribution of the measured flux
ratio for the initial 62 AGN sample.}
\includegraphics[height=5cm]{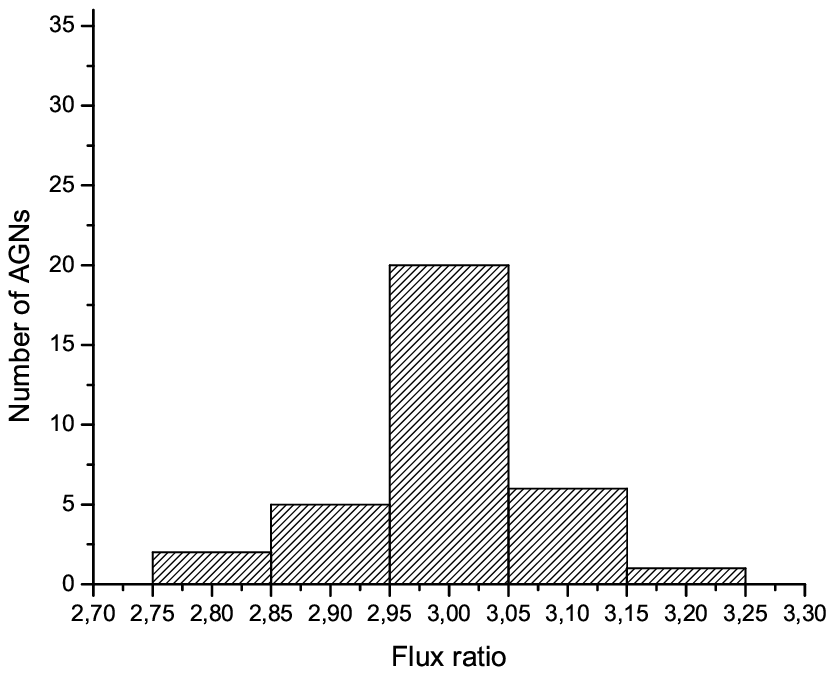}
\caption{Histogram showing the distribution of the measured flux
ratio for the final 34 AGN sample.}
\end{figure}

\subsection{Our measurements}

 The results of our measurements for these three mentioned samples
are given in Table 1 and in Figs. 4 and 5. As one can see from Table
1 and Figs. 4 and 5, discarding line profiles with slightly lower
quality does not significantly influence the result.

 The obtained flux ratio of $2.993 \pm 0.014$ is in agreement with the
theoretical improvement of \citet{b16}, who obtained the intensity
ratio of 2.98. This is better than some of the earlier values
derived from photoionized gaseous planetary nebulae and HII regions
observations, which led to the previously mentioned theoretical
reworking (\citet{b15} - 3.03; \citet{b10} - 3.17; \citet{b12} -
3.00). We should note here that measurements given in \citet{b15}
and \citet{b10} were performed in the 1980s and these differences
are probably due to the observational equipment.

\section{Conclusions}

 In order to check the improved theoretical value of the intensity
ratio of the [OIII] $\lambda$$\lambda$5007, 4959 lines we measured
the corresponding flux ratio  in a sample of 62 AGN with a high
signal-to-noise ratio spectra taken from SDSS database (DR4) and
from the observations described  in \cite{b18}. Also, from existing
literature we compiled measurements of the intensity ratio in AGN
spectra obtained as a by-product. On the basis of our investigation
we give the following conclusions:

i)  The flux ratio ($2.993\pm0.014$) obtained from our measurements
is in good agreement with the theoretical improvement obtained by
\citet{b16}, who derived an intensity ratio of 2.98. Our
observational result supports the introduction of theoretical
relativistic corrections to the magnetic dipole operator in the
calculation of the corresponding line intensity ratio.

ii) Our measured flux ratio obtained from AGN spectra is in better
agreement with the theoretical one compared to the same ratio
measured from photoionized gaseous planetary nebulae and HII regions
spectra. This is probably not caused by different physics, but by
technological advances and the better instrumentation used in
observations of AGN spectra.

iii) Despite the fact that the [OIII] lines in spectra of AGN may
have very complex line profiles, they can be used to check
sophisticated theoretical calculations. Also, we have demonstrated
that, with the development of instrumentation, the flux ratio of the
[OIII] $\lambda\lambda$4959, 5007 emission lines in AGN spectra (not
only in photoionized gaseous nebulae and HII regions spectra) may be
used to test observationally the accuracy of theoretical
calculations and to check the linearity of the detectors.

\section*{Acknowledgments}

This work is a part of the projects 146001 ``Influence of
collisional processes on astrophysical plasma line shapes'' and
146002 ``Astrophysical Spectroscopy of Extragalactic Objects''
supported by the Ministry of Science and Environment Protection of
Serbia.

\label{lastpage}

\end{document}